# Trapping of quantum particles and light beams by switchable potential wells

Eduard Sonkin[1], Boris A. Malomed[1], Er'el Granot[2] and Avi Marchewka[2]

[1] Department of Physical Electronics, School of Electrical Engineering, Tel Aviv University, Tel Aviv 69978, Israel
[2] Department of Electrical and Electronics Engineering, University Center of Samaria, Ariel 44837, Israel

## Abstract

We consider basic dynamical effects in settings based on a pair of local potential traps that may be effectively switched on and off, or suddenly displaced, by means of appropriate control mechanisms, such as the scanning tunneling microscopy (STM) or photo-switchable quantum dots. The same models, based on the linear Schrödinger equation with time-dependent trapping potentials, apply to the description of optical planar systems designed for the switching of trapped light beams. The analysis is carried out in the analytical form, using exact solutions of the Schrödinger equation. The first dynamical problem considered in this work is the retention of a particle released from a trap which was suddenly turned off, while another local trap was switched on at a distance – immediately or with a delay. In this case, we demonstrate that the maximum of the retention rate is achieved at a specific finite value of the strength of the new trap, and at a *finite value* of the temporal delay, depending on the distance between the two traps. Another ptoblem is retrapping of the bound particle when the addition of the second trap transforms the single-well setting into a double-well potential (DWP). In that case, we find probabilities for the retrapping into the ground or first excited state of the DWP. We also analyze effects entailed by the application of a kick to a bound particle, the most interesting one being a kick-induced transition between the DWP's ground and excited states. In the latter case, the largest transition probability is achieved at particular strength of the kick.

# 1. Introduction

Advances achieved in the development of the scanning tunneling microscopy (STM) have made it feasible to image and manipulate individual atoms [1,2,3]. One of remarkable results that were produced by this technology is controllable atom hopping [4,5,6]: the atom disappears at its original location and reappears at a new position. While this apparently instantaneous translation mode agrees with the principles of nonrelativistic quantum mechanics, it raises a question – how far the atom can hop as a whole, without losing electrons from its outer shell. This question is, obviously, especially relevant to Rydberg atoms (see, e.g., Ref. [7]).

The problem of maintaining the wholeness of the suddenly displaced atom against the ionization may be reduced, in the simplest form, to the consideration of a particle trapped in a potential well which instantaneously hops to a new position. In this approximation, the particle represents the loosely bound electron, while the inner part of the atom is represented by the hopping well. Then, one can calculate the probability for the initially trapped particle to remain trapped in the suddenly displaced potential.

The consideration of the same problem is also relevant in a different physical setting, when the particle is realized as the atom, while the potential well represents either the STM needle or optical tweezers, that were developed, on the basis of the laser technologies, as a tool for the trapping and release of cold atoms [8,9,10]. In this case, the atom is originally trapped at a certain location by the needle or the laser beam. Then, one switches off the voltage applied to the original needle, simultaneously applying the voltage to a needle pointing at another position. In the case of the optical trap, the original beam may be shut while focusing another beam (that may be generated by a different laser, or outcoupled from the original one) onto the new target position.

The latter problem may be readily generalized, by switching on the second local potential with a delay after the first one was switched off. In this work, we consider the delayed switch for the case of equal trapping strengths of both wells. Another physically relevant situation is that when the second well is switched on without shutting the first one, i.e., a sudden transition from the single-well setting to the double-well potential (DWP). In the latter case, we consider the symmetric setting, with equal strengths of both trapping potentials.

It is relevant to mention that dynamical effects for a quantum particle trapped in a decaying or moving potential well were recently considered in Refs. [11] and [12]. The transfer of a trapped wave packet (actually, a soliton) was also studied in terms of the nonlinear Gross-Pitaevskii equation [13], with an application to manipulations of Bose-Einstein condensates by means of laser beams [14]. The difference in the present work is that we address the dynamical situations related to the instantaneous shift of the well. For this purpose, we introduce a simple one-dimensional model, which emulates these scenarios representing the two local potentials by delta-functions (generally speaking, with different strengths). The initial configuration is the bound state in a localized potential well, the model allowing us to investigate all the dynamical effects in an analytical form. In particular, a noteworthy result is that the largest probability for the retention of the particle by the instantaneously hopping well is attained at a particular finite value of its depth (which depends on the hopping distance), rather than monotonously increasing with the depth.

As concerns the situation with the sudden switch from the single-well potential to the DWP, the symmetric set of two delta-functional traps may support an odd bound state, in addition to the even ground state, if the distance between the traps is not too small. Accordingly, in this case the switch may result in the retrapping of the originally trapped particle into either of the two bound states (even or odd), both probabilities being reported in this work.

Additionally to the above-mentioned dynamical problems, we also consider the sudden application of a kick to the trapped particle, in both cases of the single- and double-well traps. In that case, we calculate the probability of releasing the particle, as well as of the kick-induced transition between the even and odd bound states in the DWP.

The dynamical settings outlined above are also relevant for the consideration of electrons trapped by photo-switchable quantum dots, see, e.g., Ref. [15]. In particular, the settings may be used for the controllable transfer of electrons in arrays of quantum dots.

The rest of the paper is structured as follows. In Section 2 we formulate the model, in terms of local traps for the particle induced by focused laser beams. In fact, exactly the same mathematical model, based on the corresponding Schrödinger equation, applies to the description of a completely different physical system, *viz.*, transmission of optical beams in waveguiding channels, see, e.g., Ref. [16]. In this connection, we also discuss realizations and potential applications of the models, considered in the present work, in terms of the guided optical propagation. In Section 3 we report main analytical results obtained for the model based on the instantaneous or delayed switch between two potential wells. The switching from the single well to the DWP is considered in Section 4. Section 5 is dealing with the application of the kick to the bound particle, and Section 6 concludes the paper.

## 2. The model

For the quantum particle trapped in the potential induced by a focused laser beam, the setting assuming the sudden shift of the trap by finite distance $l$ is schematically presented in Fig. 1. If the particle's de Broglie wavelength is essentially larger than the width of the potential well, both the original and displaced traps may be approximated by the Fermi's potential in the form of the delta-function [11,12,17]. Thus, the model is based on the Schrödinger equation for wave function $\psi$, written in the scaled form:

$$i\frac{\partial \psi}{\partial x} = -\frac{\partial^2 \psi}{\partial x^2} + V(x,t)\psi \tag{1}$$

with the potential represented by the combination of two delta-functions – generally speaking, with different strengths, one scaled to be 2 and the other one being $2\mu$:

$$V(x,t) = \begin{cases} -2\delta(x+l) & \text{at } t \leq 0, \\ -2\mu\,\delta(x) & \text{at } t > 0, \end{cases} \tag{2}$$

The generalization of this potential for the switch delayed by time $\tau$ corresponds to

$$V(x,t) = \begin{cases} -2\delta(x+l) & at\ t \leq 0, \\ 0 & at\ 0 < t \leq \tau, \\ -2\delta(x) & at\ t > \tau, \end{cases} \tag{3}$$

where we assume the symmetric configuration, with $\mu = 1$. The other scenario that will be considered below, with the switching from the single-well trap to the DWP, is described by the following potential:

$$V(x,t) = \begin{cases} -2\delta(x+l) & at\ t \leq 0, \\ -2\delta(x) - 2\delta(x+l) & at\ t > 0, \end{cases} \tag{4}$$

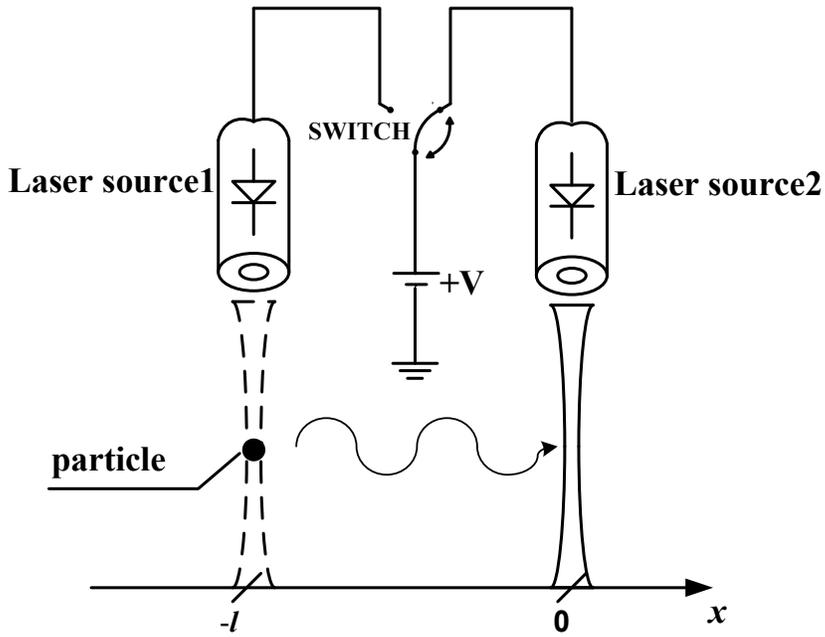

FIG. 1. A sketch of the system, realized in terms of optical beams trapping the quantum particle. Abruptly switching the pump from the left laser source to the right one implies the instantaneous displacement of the potential well by distance *l*.

The initial state, $\psi_{in}(x,t)$, is taken as the single bound state, with energy $E = -1$, which is supported by the delta-functional potential set at $x = -l$, i.e.,

$$\psi_{in}(x,t) = \exp(-|x+l| + it) \tag{5}$$

(the norm of this wave function is 1). In the case of potential (2), the normalized wave function of the finite bound state (with energy $E = -\mu^2$), shifted to point $x = 0$, is

$$\psi_{fin}(x,t) = \sqrt{\mu}\exp(-\mu|x| + i\mu^2 t) \tag{6}$$

As said above, the models may also find physically relevant interpretations in terms of guided-wave optics, in addition to the applications to quantum particles, Indeed, Eq. (1) with *t* replaced by the propagation distance, *z*, is the standard model for the paraxial transmission of light beams in planar waveguides (see, e.g., Ref. [16]), with potential $V(x,z)$ representing an effective guiding structure in the plane. In that context, all the particular potentials introduced above, *viz.*, those represented by Eqs. (2), (3), and (4), may find their applications to optics, in terms of the switching or splitting of light signals between different channels, see Ref. [18]. Accordingly, the results reported below may be interpreted in terms of the guided waves in linear optics.

# 3. Analytical results for the single-well potentials

## 3.1 The instantaneous shift of the potential well

The probability for the particle, which was originally trapped in bound state (5), to get retrapped into state (6) after the sudden shift of the potential well and the change of its strength from 1 to $\mu$, can be calculated in an obvious way, as the square of the respective overlap integral:

$$P(\mu) \equiv |A(\mu)|^2 = \frac{4\mu(\mu e^{-l} - e^{-\mu l})^2}{(\mu^2 - 1)^2}, \tag{7}$$

where $A(\mu) \equiv \int_{-\infty}^{+\infty} \psi_{in}^{*}(x)\psi_{fin}(x)dx$, with the asterisk standing for the complex conjugation. Expression (7) is not defined for the case of the symmetric configuration, with $\mu = 1$. However, it is easy to resolve this case, taking the limit of $\mu - 1 \to 0$:

$$P(\mu = 1) = (1+l)^2 e^{-2l}. \tag{8}$$

Note that Eqs. (7) and (8) yields $P < 1$ for any $l > 0$. The probability given by these expressions is plotted in Fig. 2 versus $\mu$, for several fixed values of distance $l$ between the initial and final positions of the trap. A noteworthy feature of the plots is that the maximum of the retention probability is attained at a finite value of $\mu$, and the probability vanishes at $\mu \to \infty$, i.e., an extremely strong shifted potential trap is inefficient in retaining the particle in the bound state. Accordingly, a value $\mu_{max}$, at which the maximum of the probability ($P_{max}$) is found by means of a straightforward analysis of expression (7), and $P_{max}$ itself, are plotted in Fig. 3 versus $l$. In particular, in the limit of $l >> 1$ one can easily obtain the following analytical approximations from Eq. (7):

$$P(\mu) \approx 4\mu \exp(-2\mu l), \ \mu_{max} \approx (2l)^{-1}, \ P_{max} \approx (2/e)l^{-1}. \tag{9}$$

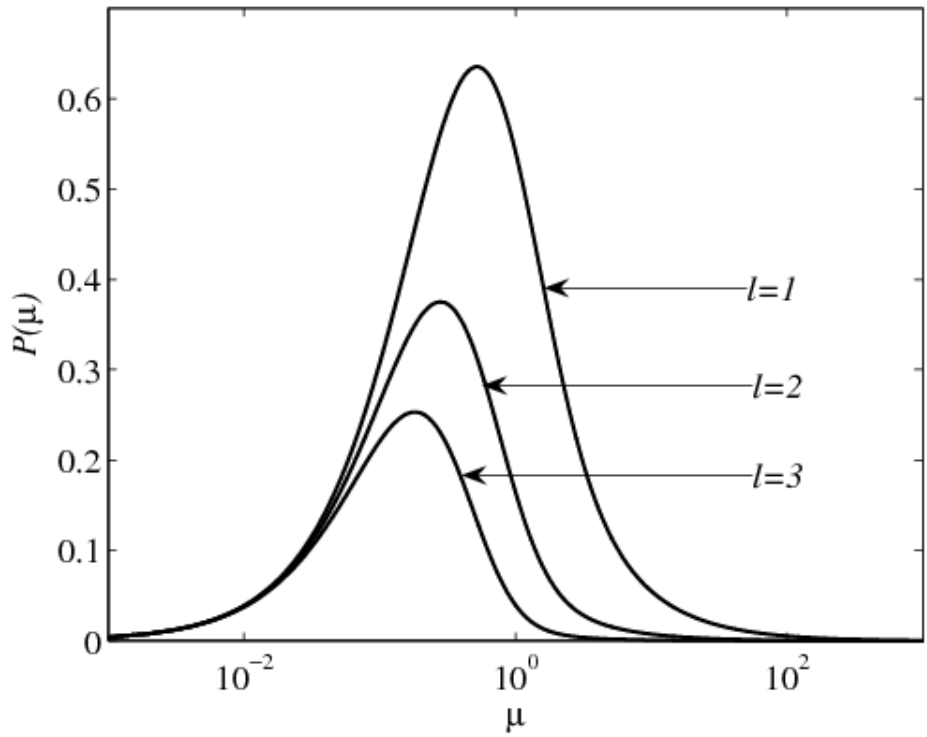

FIG. 2. The probability of the particle to remain trapped after the instantaneous shift of the potential well by distance *l* is shown versus relative strength $\mu$ of the shifted well, on the logarithmic scale.

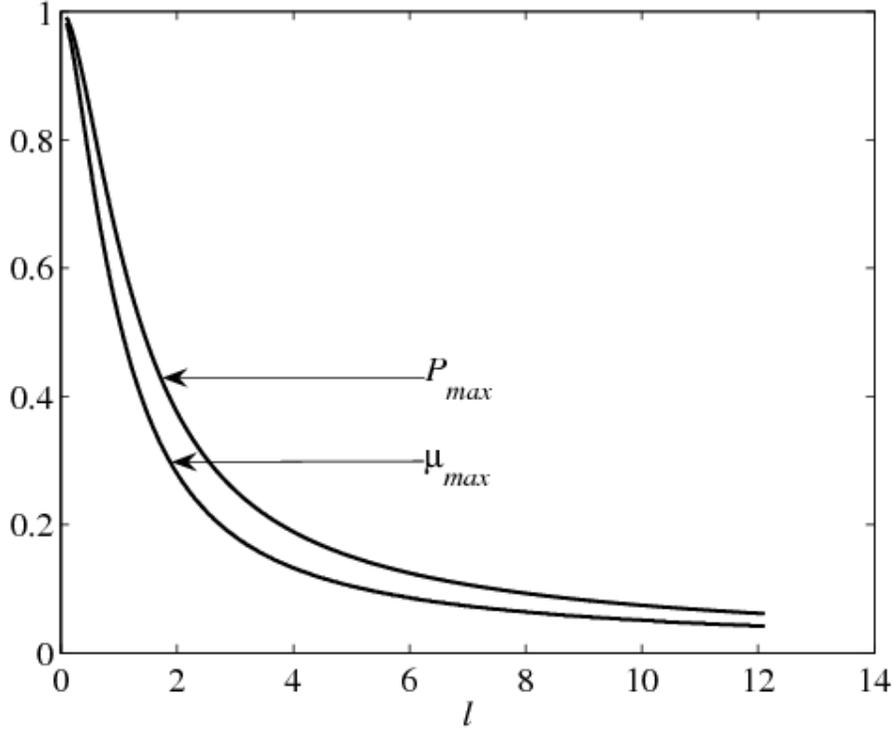

FIG. 3. The relative strength of the shifted trap ($\mu_{max}$), which provides for the maximum of the retention probability, $P_{max}$, and the maximum itself, are shown as functions of the shift, $l$.

In particular, the asymptotic relation $\mu_{max} \approx (2l)^{-1}$ in Eq. (9) may be understood as indicating that the strongest retention is provided by matching of the two momenta which are determined, via the Heisenberg's uncertainty relation, by the displacement of the potential well, $p_l \sim 1/l$, and by the localization length of the final bound state, $p_\mu \sim \mu$.

A solution for the temporal evolution of the wave function in the course of the retrapping is available too. The time-dependent solution to Schrödinger equation (1) with potential (2) and initial conditions given by Eq. (5) at $t = 0$ can be written, at $t > 0$, by means of the Green's function (kernel) $K(x, x'; t)$:

$$\psi(x,t) = \int_{-\infty}^{+\infty} K(x, x'; t) \psi_{in}(x', t=0) dx'. \qquad (10)$$

In the presence of the final potential, $V(x) = -2\mu\delta(x)$, the relevant Green's function is [19]

$$K(x, x'; t) = K_{free}(x, x'; t) + \frac{\mu}{2} e^{-\mu(|x|+|x'|-i\mu t)} \operatorname{erfc}\left( \frac{|x|+|x'|-2i\mu t}{2\sqrt{it}} \right), \qquad (11)$$

where erfc($x$) is the standard complementary error function, and the free-space kernel is

$$K_{\text{free}}(x, x';t) \equiv \frac{1}{2\sqrt{i\pi t}} \exp\left(i\frac{(x-x')^2}{4t}\right). \tag{12}$$

The solution given by Eqs. (10)-(12) can be evaluated explicitly, although the eventual form is cumbersome:

$$\psi_\mu(x,t) = M(x+l,-i,2t) + M(-x-l,-i,2t)$$
$$+ \frac{2\mu}{1-\mu^2} M(|x|+l,i\mu,2t) - \frac{2\mu^2 e^{-l}}{1-\mu^2} M(|x|,i\mu,2t) - \frac{\mu}{1+\mu} M(|x|+l,-i,2t)$$
$$- \frac{\mu}{1-\mu} M(|x|+l,i,2t) + \frac{\mu e^{-l}}{1-\mu} M(|x|,i,2t) - \frac{\mu e^{-l}}{1+\mu} M(|x|,-i,2t), \tag{13}$$

where the definition of the so-called *Moshinsky function* [17,20,21] has been used,

$$M(x,k,t) \equiv \frac{1}{2} \exp\left(i\frac{x^2}{2t} - z^2\right) \text{erfc}(iz), \quad z \equiv \frac{1+i}{2}\sqrt{t}\left(k - \frac{x}{t}\right), \tag{14}$$

$k$ being a complex parameter. Note that for $\mu = 0$, i.e., after the abrupt elimination of the original potential without switching the new potential on, the explicit solution (13) simplifies in the free space:

$$\psi_0(x,t) = \frac{1}{2} e^{it} \left\{ e^{x+l} \text{erfc}\left(\sqrt{it} + \frac{x+l}{2\sqrt{it}}\right) + e^{-x-l} \text{erfc}\left(\sqrt{it} - \frac{x+l}{2\sqrt{it}}\right) \right\}. \tag{15}$$

The asymptotic form of the free-space solution (15) for $t \to \infty$ may be further simplified, making use of the known asymptotic approximation, $\text{erfc}(z) \approx \left(\sqrt{\pi} z\right)^{-1} e^{-z^2}$ [22]:

$$\psi_0(x,t) \cong \frac{t^{3/2}}{\sqrt{i\pi}} \frac{e^{\frac{i(x+l)^2}{4t}}}{t^2 + \left(\frac{x+l}{2}\right)^2}. \tag{16}$$

In the case of the symmetric pair of the delta-functions, with $\mu = 1$, general solution (13) looks singular. After some algebra, one can resolve the singularity, cf. the transition from Eq. (7) to Eq. (8). In that case, the explicit solution is

$$\psi_{\mu=1}(x,t) = \frac{1}{2}e^{it}\left\{e^{x+l}\,\text{erfc}\left(\sqrt{it} + \frac{x+l}{2\sqrt{it}}\right) + e^{-x-l}\,\text{erfc}\left(\sqrt{it} - \frac{x+l}{2\sqrt{it}}\right)\right\}$$

$$+\left(\frac{1}{2} + |x| + l - 2it\right)\frac{1}{2}e^{it-|x|-l}\,\text{erfc}\left(-\sqrt{it} + \frac{|x|+l}{2\sqrt{it}}\right) - \frac{1}{4}e^{it+|x|+l}\,\text{erfc}\left(\sqrt{it} + \frac{|x|+l}{2\sqrt{it}}\right)$$

$$+e^{-l}\left(\frac{1}{2} - |x| + 2it\right)\frac{1}{2}e^{it-|x|}\,\text{erfc}\left(-\sqrt{it} + \frac{|x|}{2\sqrt{it}}\right) - \frac{e^{-l}}{4}e^{it+|x|}\,\text{erfc}\left(\sqrt{it} + \frac{|x|}{2\sqrt{it}}\right)$$

$$-\sqrt{\frac{it}{\pi}}e^{-\frac{(|x|+l)^2}{4it}} + \sqrt{\frac{it}{\pi}}e^{-\frac{x^2}{4it}-l}. \tag{17}$$

Using the explicit result (13), one can produce probability density distributions which illustrate the evolution of the wave function in the course of the retrapping, see typical examples of intermediate and nearly-final configurations in Fig. 4. In addition, the evolution of the probability density of the exact solution at $x = 0$ (the point at which the new potential trap was created) is plotted, for the same case, in Fig. 5.

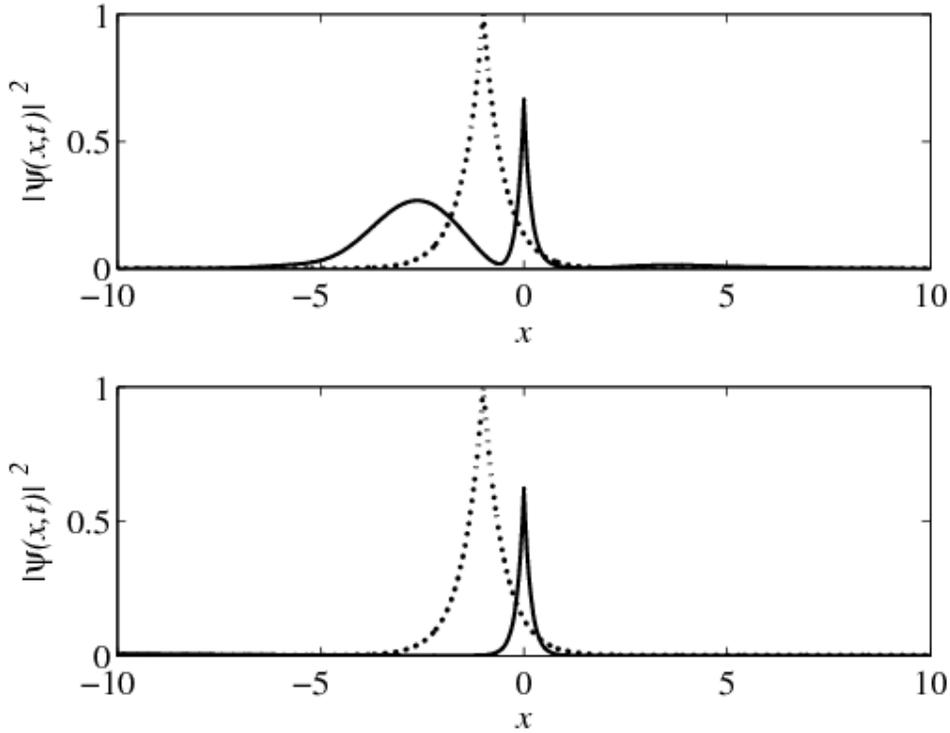

FIG. 4 The top and bottom panels display the distribution of the probability density in space at two different moments of times, $t = 0.07$ and $15$, respectively, in the course of the switching with $\mu = 3$, $l = 1$, and $\tau = 0$ (zero delay). The dotted line represents the initial state $|\psi(x,t=0)|^2$, and the solid curve depicts $|\psi(x,t)|^2$. The final retention probability is given by Eq. (7), $P(\mu = 3, l = 1) \approx 0.21$.

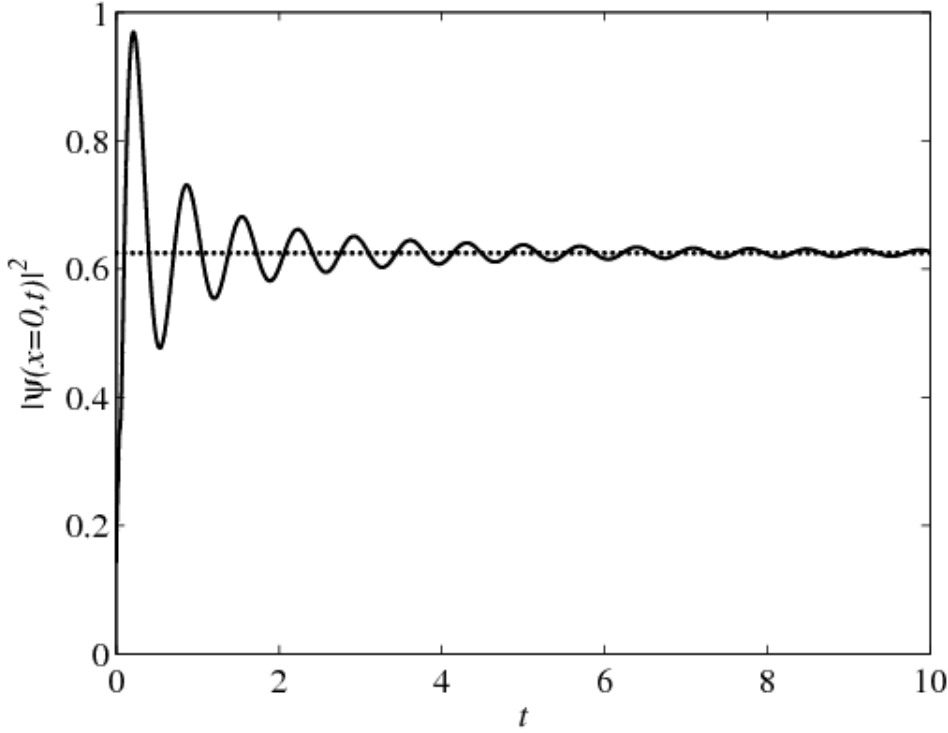

FIG. 5. The temporal evolution of the probability density at $x=0$, for $\mu=3$ and $l=1$ (the oscillating curve), in the case of the switching with zero delay. The horizontal (dotted) line stands for the final value, $|\psi(0,t\to\infty)|^2$.

In the case of the symmetric pair of the delta-functions, with $\mu=1$, Eq. (17) makes it possible to find a simple asymptotic expression for the wave function at $x=0$, $\psi_{\mu=1}(0,t\to\infty)=e^{it-l}(1+l)$. As follows from Eq. (6), this result complies with the retention probability (7) obtained above for $\mu=1$.

### 3.2. The delayed retrapping of the particle

In the model based on the time-dependent potential in the form of Eq. (3), one can use the free-space solution (15) to find the wave function at the moment of $t=\tau$, and then the delayed-retrapping probability can be found via the following overlap integral:

$$P=|A(\tau)|^2, \text{ where } A(\tau)\equiv\int_{-\infty}^{+\infty}\psi_{\text{fin}}^*(x,0)\psi(x,\tau)dx, \tag{18}$$

where expression (6) for $\psi_{fin}(x,0)$ is taken with $\mu = 1$. Using Eq. (15), one can explicitly evaluate this integral:

$$A(\tau) = -2i\tau e^{-l}[M(0,-i,2\tau) + M(0,i,2\tau)] + (1 - 2i\tau - l)M(l,-i,2\tau)$$
$$- (1 - 2i\tau + l)M(l,i,2\tau) + 2\sqrt{\frac{i\tau}{\pi}} \exp\left(\frac{il^2}{4\tau}\right) + (1+l)\exp(i\tau - l). \quad (19)$$

The probability given by Eqs. (18) and (19) is plotted versus delay time $\tau$ in Fig.6, at several fixed values of shift $l$. A noteworthy feature of this dependence is the existence of well-pronounced maxima of the retrapping probability at finite values of $\tau$, provided that $l$ is not too small ($l > 1$). Moreover, the finite delay gives rise to the *strong enhancement* of the retention rate of the particle at large values of the shift, such as $l = 4$ and 5.

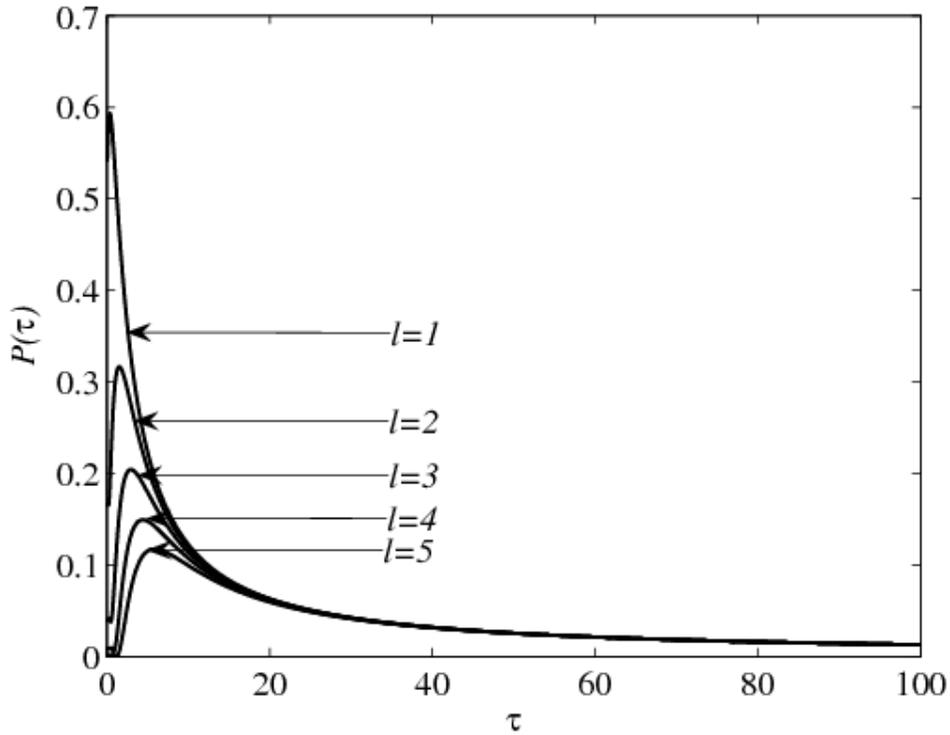

FIG. 6. The probability of the delayed retrapping of the quantum particle as a function of the delay time, at fixed values of displacement $l$.

# 4. The double-well potential

Proceeding to the consideration of the model based on the switching from the single well to the DWP, as per Eq. (4), we first consider static bound states supported by the symmetric DWP. It is easy to find the even stationary wave function of the ground state, $\phi_{\text{even}}(x)$, and its counterpart for the first excited state, $\phi_{\text{odd}}(x)$ (provided that the latter one exists) [23]:

$$\phi_{\text{even}}(x) = C_{\text{even}} \begin{cases} e^{\alpha_{\text{even}}(x+l)} & \text{at} \quad x \leq -l, \\ \dfrac{\cosh(\alpha_{\text{even}}(x+l/2))}{\cosh(\alpha_{\text{even}}l/2)} & \text{at} \quad -l < x \leq 0, \\ e^{-\alpha_{\text{even}}x} & \text{at} \quad x > 0; \end{cases} \qquad (20)$$

$$\phi_{\text{odd}}(x) = C_{\text{odd}} \begin{cases} -e^{\alpha_{\text{odd}}(x+l)} & \text{at} \quad x \leq -l, \\ \dfrac{\sinh(\alpha_{\text{odd}}(x+l/2))}{\sinh(\alpha_{\text{odd}}l/2)} & \text{at} \quad -l < x \leq 0, \\ e^{-\alpha_{\text{odd}}x} & \text{at} \quad x > 0, \end{cases} \qquad (21)$$

where $C_{\text{even,odd}}$ are the respective normalization constants, and $\alpha_{\text{even,odd}}$ are related to the separation between the two wells, $l$, by transcendental equations following from the integration of Eq. (1) in infinitesimal vicinities of points $x = -l$ and $x = 0$:

$$\alpha_{\text{even}}[1 + \tanh(\alpha_{\text{even}}l/2)] = 2, \quad \alpha_{\text{odd}}\coth(\alpha_{\text{odd}}l/2) = 2 - \alpha_{\text{odd}}. \qquad (22)$$

As follows from Eqs. (22), the even wave function represents the single solution at $l < 1$, while at $l > 1$ the odd wave function exists too. Finally, the energies of the ground state and first excited states can be written as

$$E_{\text{ground}} = -\alpha_{\text{even}}^2, \quad E_{\text{excited}} = -\alpha_{\text{odd}}^2. \qquad (23)$$

Absolute values of energies (23) are displayed as functions of $l$ in Fig. 7.

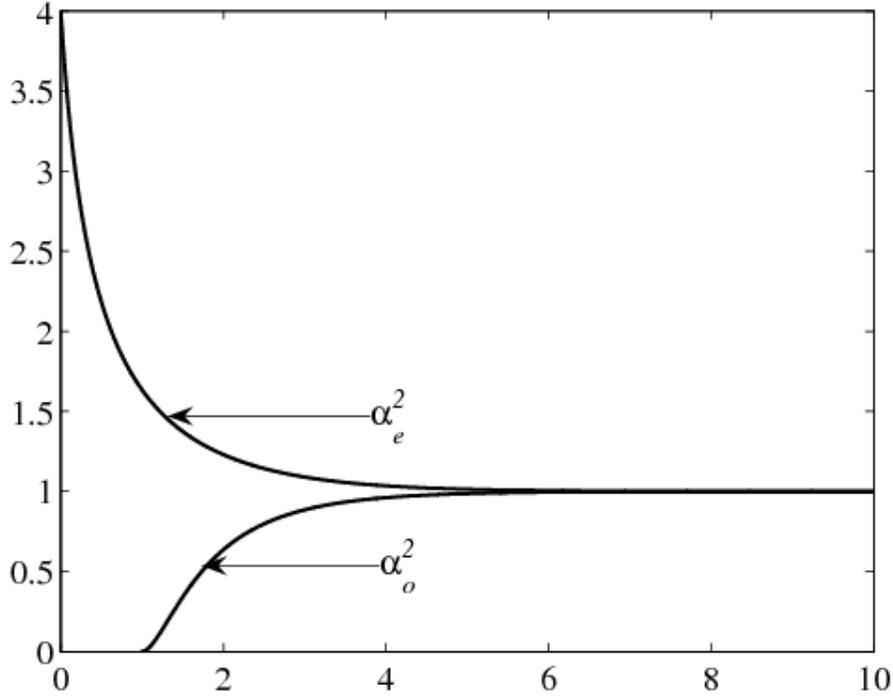

FIG. 7. The absolute values of the energies of the ground state and first excited states (alias even and odd ones, "e" and "o", respectively) in the double-well potential, found as per Eqs. (22) and (23). The odd wave function exists at $l > 1$.

Getting back to the model with the full time-dependent potential (4), the probabilities of retrapping the particle from the initial state (5) into the even and odd bound modes, which are supported by the DWP, are given, respectively, by the following expressions:

$$P_{\text{even,odd}}(l) = C^2_{\text{even,odd}} \left| \int_{-\infty}^{+\infty} \psi^*_{\text{in}}(x, t=0) \phi_{\text{even,odd}}(x) dx \right|^2. \qquad (24)$$

These probabilities are shown, as functions of $l$, in Fig. 8. Note that the two probabilities become practically equal at $l > 6$.

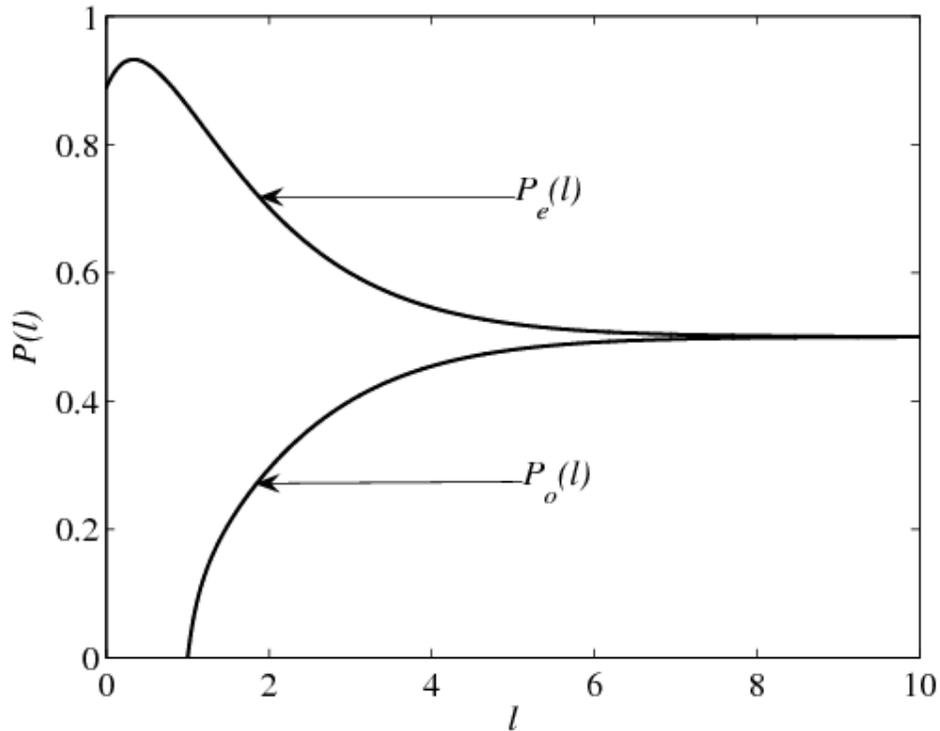

FIG. 8. The probabilities for the retrapping of the original state (5) into the ground (even) state ("e") and the first excited (odd) state ("o"), evaluated as per Eqs. (20)-(22) and (24) in the model based on Eq. (4). Recall that the excited state exists at $l >1$.

## 5. The relaxation of the kicked trapped particle

Dynamical effects of another type are induced by the application of a *kick* to a trapped particle, i.e., imparting momentum $k$ (or kinetic energy $k^2$) to it, instantaneously multiplying the stationary wave function by $\exp(ikx)$. A natural realization of the kick is provided by the situation when the trapping potential is abruptly set into motion with velocity $v = -2k$. On the other hand, if the above models are interpreted in terms of the guided-wave transmission in optics, as outlined above, the application of the kick implies a sudden change of the propagation direction in the planar waveguide.

A natural question is the probability of the retention of the kicked particle in the bound state. For instance, if the kick is applied to the stationary wave function (5), the holding probability can be readily calculated as

$$P(k^2) = |A(k)|^2, \quad A(k) \equiv \int_{-\infty}^{+\infty} e^{ikx} |\psi_{in}(x, t=0)|^2 dx = \frac{4}{4+k^2} . \tag{25}$$

In the DWP setting, an interesting possibility is the kick-induced transition between the even ground state and its odd excited counterpart. The probability of this transition (in either direction) can be evaluated as per the following expression, cf. Eq. (25):

$$P_{trans}(k^2) = |A(k)|^2, \quad A(k) \equiv \int_{-\infty}^{+\infty} e^{ikx} \phi_{even}^*(x) \phi_{odd}(x) dx, \tag{26}$$

where $\phi_{even}(x)$ and $\phi_{odd}(x)$ are the stationary wave functions given by Eqs. (20) and (21). The result of the numerical calculation of this probability is presented in Fig. 9, which features a noteworthy fact: the transition probability attains a maximum, $P_{max}$, at a finite value of the kinetic energy lent to the originally quiescent particle. The latter value, along with the respective maximum $P_{max}$, are shown as functions of the DWP size, $l$, in Fig. 10. In the same figure, the energy difference between the even and odd states is shown too, for comparison.

In addition to the presence of the maxima, Fig. 9 demonstrates vanishing of the transition probability at particular values of $k$. The existence of both the maxima and zeros of $P_{trans}(k^2)$ can be easily explained. Indeed, the product $\phi_{even}(x)\phi_{odd}(x)$ takes opposite values at points $x = -l$ and $x = 0$. In the combination with the phase shift of $kl = \pi(2n-1)$ or $kl = 2\pi n$, where $n = 1, 2, 3...$, expression (26) clearly predicts, severally, maxima and zeros of $P_{trans}(k^2)$. In particular, the maxima and zeros observed in Fig. 9 are well explained by these formulas with $n = 1$ and 2.

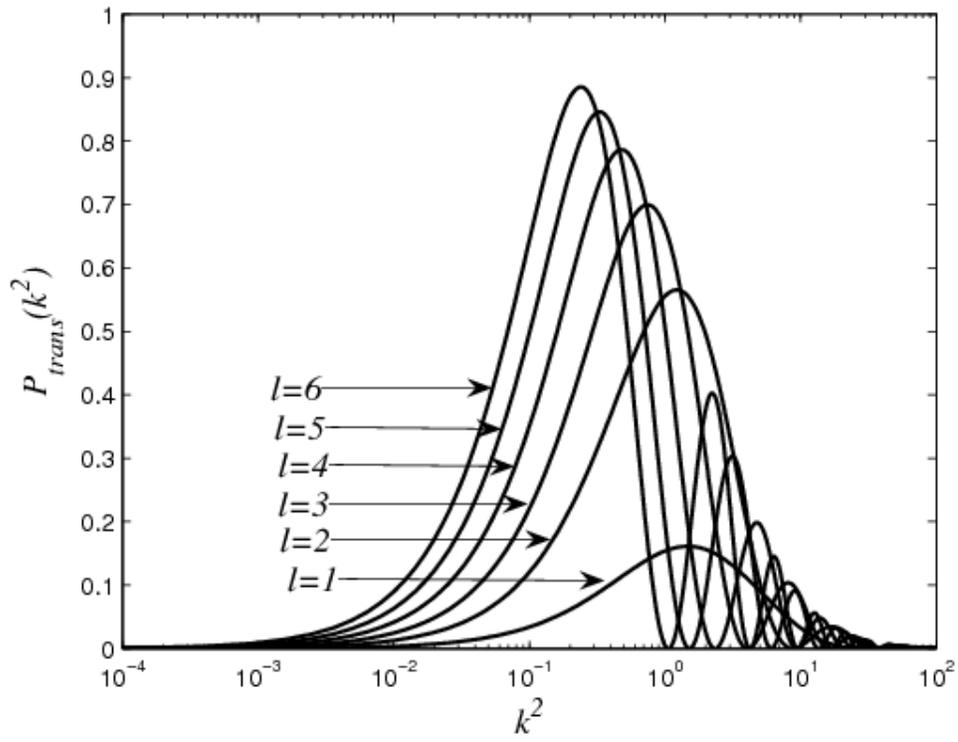

Fig. 9: The probability for the transition of the particle trapped in the double-well potential between the ground state and first excited state under the action of the kick, shown versus the kinetic energy $k^2$ imparted to the particle.

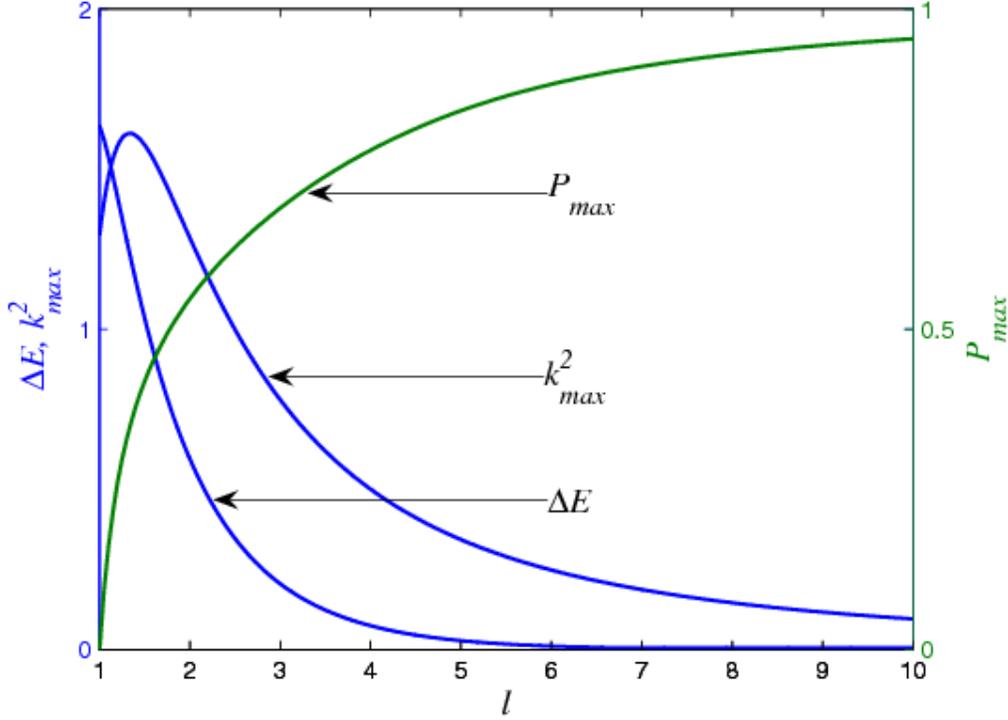

Fig. 10: (Color online) The kinetic energy, $k_{max}^2$, of the kick, which gives rise to the maximum of the probability of the transition between the ground state and first excited state in the DWP, is plotted versus the distance between the two potential wells. The maximal probability, $P_{max}$, is shown too, as well as the energy difference, $\Delta E$, between the two bound states [see Eq. (23)].

## 6. Conclusions

In this paper we have addressed several dynamical effects expected in switchable configurations of local traps for quantum particles. These include retention of the particle released from a trap which was shut down, followed by switching on another trap at a distance from the original one – immediately, or with a finite delay; retrapping of the particle into the ground state, or its first-excited-state counterpart, in the case when the addition of the extra potential well suddenly turns the single-well potential into the DWP (double-well potential); and effects produced by the application of the kick to a trapped particle, such as the transition between the ground and excited states in the DWP. The results have been obtained in the analytical form and were given physical explanations. In addition to atoms or electrons trapped in scanning-tunneling-microscopy settings, or captured by switchable qauntum dots, the same models, and the results produced by the analysis, apply to the problem of switching of light beams in planar waveguiding structures.

A challenging generalization of the dynamical problems considered in this work may be extending them to nonlinear systems, including those with the uniform nonlinearity [24] and models with the trapping provided by *nonlinear* potential wells, such as those considered in Refs. [23] and [25]. Another potentially interesting extension may be expected in two-dimensional settings with switchable potential wells.